\documentclass[preprint,showpacs,preprintnumbers,amsmath,amssymb,a4paper,superscriptaddress]{revtex4}
\usepackage{graphicx}% Include figure files
\usepackage{dcolumn}% Align table columns on decimal point
\usepackage{bm}% bold math

\begin{document}

%\preprint{APS}

\title{Quantum Scattering in Strong Cylindrical Confinement}

\author{Ji il Kim}
\affiliation{Physikalisches Institut, Universit\"{a}t Heidelberg, Philosophenweg 12, 69120
Heidelberg, Germany}
\author{J\"{o}rg Schmiedmayer}
\affiliation{Physikalisches Institut, Universit\"{a}t Heidelberg, Philosophenweg 12, 69120
Heidelberg, Germany}
\author{Peter Schmelcher}
\affiliation{Physikalisches Institut, Universit\"{a}t Heidelberg, Philosophenweg 12, 69120
Heidelberg, Germany}
\affiliation{Theoretische Chemie, Institut f\"{u}r Physikalische Chemie, Universit\"{a}t
Heidelberg, Im Neuenheimer Feld 229, 69120 Heidelberg, Germany}
%\altaffiliation[Also at]{Physics Department, XYZ University.}%Lines break automatically or can be forced with \\
%\email{author@institution.edu}

%\affiliation{

\date{24 Mar. 2005}

\begin{abstract}
A Green's function formalism to analyze the scattering properties in confined geometries
is developed. This includes scattering from a central field inside the guide created
e.g. by impurities. For atomic collisions our approach applies to the case of parabolic
confinement and, with certain restrictions, also to an anharmonic one. The coupling
between the angular momentum phase shifts $\delta_l$ of a spherically symmetric scattering
potential $V(r)$ due to the cylindrical confinement is analysed. Under these general
conditions, a broad range of scattering energies covering many transversal excitations is
considered and changes to the bound states of $V(r)$ are derived. For collisions between
identical atoms, the boson-fermion and fermion-boson mappings are demonstrated. 
\end{abstract}

\maketitle

%\section{\label{sec:level1}First-level heading:\protect\\ The line
%break was forced \lowercase{via} \textbackslash\textbackslash}

Implementing atom-optical devices often requires a strong confinement for all except one
degree of freedom~\cite[and Refs therein]{intro}. Examples of physical situations where a
strong confinement is needed are guided matter-wave interferometers~\cite{intro}, one
dimensional optical lattices~\cite{morsch2002a}, cold gases in very elongated traps for
studies of superfluidity~\cite{cataliotti2003a}, the Tonks-Girardeau
gas~\cite{paredes2004a} or phase fluctuations of quasi-condensates~\cite{dettmer2001a}. A
proper description of the dynamics of such reduced quasi-1D systems should account for the
nature of the discrete transverse states. Therefore one needs to deduce the effective 1D
interaction between the remaining longitudinal degrees of freedom from the real 3D
free-space interaction potential $V(r)$. Collisions under confinement different from the
1D case are treated in~\cite{petrov2000b,diffg}. Apart from ultracold {\em atom-atom 
collisions}, scattering in confined geometries also occurs in various physical situations
such as scattering of guided atomic matter waves or of guided electromagnetic and
accoustic waves~\cite{olsson1981} from {\em obstacles inside a guide}, e.g., (heavy)
impurity atoms or material defects, respectively. The latter is of importance for the
propagation of radiation or sound within transmission lines or resonators.  

As for atom-atom collisions, resonant quasi-1D scattering in the transverse ground state 
of the guide (single mode regime) was first considered for bosons in an harmonic guide 
employing for the interaction potential $V(r)$ a delta-like zero-range
approximation~\cite{olshanii1998a}. Numerical simulations~\cite{bergeman2003} confirmed
for certain finite range potentials $V(r)$ the existence of the so-called confinement
induced resonance (CIR) originally predicted in~\cite{olshanii1998a}. A further
investigation of the CIR is provided in~\cite{granger2004a} dealing for the first time
with a general finite-range $V(r)$ for both bosons and fermions under harmonic
confinement. Effects of the non-parabolicity of the confinement are considered
in~\cite{peano2004a}, with focus on the center of mass dynamics and employing a zero-range
approximation for the interaction $V(r)\,$. 

The present work extends the above approaches and gives an alternative and complementary
description of scattering under confinement, treating both the cases of collisions and of 
scattering by fixed obstacles. We develop a general formalism based on the Green's
functions that allows us to express the scattering properties in confined geometries in
terms of the phase-shifts $\delta_l$ of free-space scattering. The coupling between these
phase-shifts is explicitly taken into account. A general initial scattering state can be
treated properly, describing in particular the ``multi-channel'' regime, in the sense that
the total energy allows several transversal excited states to be effectively occupied. 

In the case of collisions where $V(r)$ is the atom-atom interaction potential, the center
of mass motion is known to separate from the relative one only for a {\em parabolic
confinement}. Our approach then provides a deeper understanding of this collision
process. On the other hand, for an {\em arbitrary confinement}, scattering processes that
can naturally be described by the formalism include, e.g., the quantum scattering of
individual cold atoms, or other equivalent systems, by a central field $V(r)$ fixed in the
center of the guide at $\bm{r}=0$. As for atom-atom scattering, the relative coordinates
$\bm{r}$ are not exactly separable from the center of mass coordinates $\bm{R}\,$ if the
confinement is no longer parabolic. Nevertheless, in such a situation of coupled center of
mass and relative motion, the formalism provides in the ultracold regime a distinct 
starting-point to account for this coupling.  

%Nevertheless, the formalism should be applicable to a restricted set of
%collisions for which the center of mass reference frame is not accelerated. This is the
%case at low temperatures and single mode regime, when the atoms are mostly stabilized
%along the symmetry axis of the confining potential before the collision takes place. Then
%the center of mass $\bm{R}$ is disturbed only by quantum fluctuations, since it is not
%acted on by the interaction $V(r)$. As the anharmonicities of the confinement are reduced
%and thereby the quantum fluctuations effects of the center of mass on the relative
%coordinates decrease, one expects an increasingly higher accuracy in the description of
%atom-atom  scattering. It turns out as a result that, in the limit of parabolic
%confinement, this approach provides a deeper understanding of the collison process under
%confinement. 

Under the above restrictions concerning atom-atom collisions, our investigation confirms  
that the CIR~\cite{olshanii1998a} is a general consequence of the dominant terms of the
scattering amplitudes. The main requirements are a large positive $s$-wave scattering
length $a$, $a\sim l_\perp$ [$l_\perp$ is the length scale of the confining 
potential $U(\rho)\,$, such that $U\approx 0$ for $\rho\ll l_\perp$, and equals the
cylinder radius for a square-well type confinement], a short-ranged scattering potential
$V(r)$, $R_V\ll l_\perp$ [$R_V$ is the range of $V(r)\,$, such that $V\approx 0$ for $r\gg
R_V$], small longitudinal momenta and small phase-shifts $\delta_l$, as described below. 
%We show that these conditions of validity are related to the current conservation. 
The resonance is accompanied by a $l=0$ bound-state of $V(r)$ strongly distorted by the
confinement $U(\rho)$ and pushed towards the continuum. This modified bound
state~\cite{bergeman2003} is shown to be a herald of the CIR. In the context of scattering
of individual guided atoms by a central field, these conclusions hold irrespective of
restrictions due to anharmonicities and imply the unambiguous strong effects of
confinement on the scattering process.

{\em Phase-Shifts}.\,  The Schr\"odinger equation for the scattering wave function
$\Psi(\bm{r})$, with $\bm{r}=(\bm{\rho},z)$, reads 
\begin{equation}
\label{diff}
\left[\nabla^2 - u(\rho) + k^2\right]\Psi(\bm{r}) = v(r)\Psi(\bm{r}),
\end{equation}
where $u(\rho)\equiv 2\mu U(\rho)/\hbar^2$, $v(r)\equiv 2\mu V(r)/\hbar^2$, and
$E=\hbar^2k^2/2\mu > 0$ is the total energy. In the case of atomic collisions, $\bm{r}$
is the relative coordinate and the relation between $U(\rho)$ and the confining potential
$U_c(\rho_i)$ of the $i$-th particle in the laboratory reference frame is given by
$U(\rho)=2\,U_c(\rho/2)\,$. Note that this relation is no longer exact for non-parabolic
$U_c$ (but provides a first uncoupled description of the relative motion, by quenching the
center of mass at the origin $\bm{R}=0$). The cylindrical boundary condition is met by
expanding the solution in the transverse eigenstates $\varphi_n(\rho)$, with energies  
$\epsilon_n\equiv \hbar^2q_n^2/2\mu$ and normalized to $\int 
dxdy\,\varphi_n(\rho)^\ast\varphi_m(\rho)=\delta_{nm}$. As a result, one obtains the
integral equation 
\begin{equation}
\label{int}
\Psi(\bm{r}) = \Psi_i(\bm{r})-\int d^3\bm{r}^\prime 
		G_c(\bm{r},\bm{r}^\prime) v(r^\prime)\Psi(\bm{r}^\prime).
\end{equation}
For a given $k$ low enough such that $k\sim 1/l_\perp$, let $n_E$ be the integer obeying
$k^2=q_{n_E}^2+k_{n_E}^2\leq q_{1+n_E}^2$. The following study includes the situations of
ground state scattering ($n_E=0$) as well as scattering in the {\em transversally excited}
modes ($n_E\geq 1$). In both cases, transverse states with $n>n_E$ can only be {\em
virtually} occupied, since $k^2<q_n^2\,$. The general initial state is 
$\Psi_i(\bm{r})=\sum_{n=0}^{n_E}b_ne^{ik_nz}\varphi_n(\rho)$ for some constants $b_n$,
with $k^2=q_n^2+k_n^2$. In Eq.(\ref{int}), 
\begin{equation}
\label{gc}
G_c(\bm{r},\bm{r}^\prime) =
\sum_{n=0}^\infty\varphi_n(\rho)\varphi_n(\rho^\prime)^\ast G_n(z-z^\prime) 
\end{equation}
is an axially symmetric Green's function and $G_n(z)=-e^{ik_n|z|}/2ik_n$ (for $n\leq n_E$) and
$G_n(z)=e^{-p_n|z|}/2p_n$ (with $k^2=q_n^2-p_n^2$, for $n\geq 1+n_E$) are 1D Green's 
functions. The excited states with quantum numbers larger than $n_E$ decrease
exponentially with increasing distance from the  scattering region. In the asymptotic
limit $|z|\rightarrow\infty$, one has for $n\leq n_E$  
\begin{subequations}
\label{c-boundary}
\begin{eqnarray}
\label{asymptotic}
\hspace{-1.5em}
\Psi(\bm{r}) &\approx& 
\sum_{n=0}^{n_E}\left[b_ne^{ik_nz} + f_n^\pm\,e^{ik_n|z|}\right]\varphi_n(\rho),
		z\rightarrow\pm\infty\, , \\
\label{f1d}
\hspace{-1em}
f_n^\pm &\equiv & \frac{1}{2ik_n}\int d\bm{r^\prime}\left[e^{\pm ik_nz^\prime}
	\varphi_n(\rho^\prime)\right]^\ast v(r^\prime)\Psi(\bm{r}^\prime),
\end{eqnarray} 
\end{subequations}
where $f_n^\pm$ is the $n$-th channel {\em effective 1D scattering amplitude} for forward
$z>0$ and backward $z<0$ scattering. 

Consider next $G_c(\bm{r},\bm{r}')$ for $r'< r \ll l_\perp$. In this region
$U(\rho)\approx 0$ and one should be able to approximate $G_c$ by the free 3D Green's
functions $G_{1,2}(\bm{r},\bm{r}^\prime)\equiv e^{\pm
ik|\bm{r}-\bm{r}^\prime|}/4\pi|\bm{r}-\bm{r}^\prime|$. Thus, we write  
\begin{subequations}
%\begin{widetext}
\begin{eqnarray}
\label{green-3D}
G_c(\bm{r},\bm{r}^\prime) & = & \frac{1}{2\pi}\int d\phi^\prime 
	\left(\gamma_+ \frac{e^{ik|\bm{r}-\bm{r}'|}}{4\pi|\bm{r}-\bm{r}'|} 
		+ \gamma_- \frac{e^{-ik|\bm{r}-\bm{r}'|}}{4\pi|\bm{r}-\bm{r}'|}\right)
%	\left[\gamma_+ G_1(\bm{r},\bm{r}') + \gamma_- G_2(\bm{r},\bm{r}')\right]
								\nonumber\\
	& & \hspace{0.5em}
		+ \hspace{0.5em} \Delta_c(\bm{r},\bm{r}^\prime) 	 \\
\label{green}
	&=&  ik\sum_l j_l(kr') 
		\left[ \gamma_+ h_l^{(1)}(kr) - \gamma_- h_l^{(2)}(kr) \right]
								\nonumber\\
	& & \hspace{5em}
		\times\frac{2l+1}{4\pi}P_l(\cos{\theta})P_l(\cos{\theta'})
								\nonumber\\
	& & \hspace{1em}
		+ \hspace{0.5em} \Delta_c(\bm{r},\bm{r}^\prime)\, ,
			\hspace{3.5em} r'< r \ll l_\perp.
\end{eqnarray}
%\end{widetext}
\end{subequations}
In Eq.(\ref{green}), we have used the well known expansion of $G_{1,2}$ in spherical 
coordinates~\cite[Prob.7.5]{morse1953}. The value of $\gamma_{\pm}$ and $\Delta_c$ can be
explicitly obtained if $U(\rho)$ is approximated by a {\em square-well} type 
confinement for $r\ll l_\perp$. Indeed, the eigenstates are then close to Bessel
functions, $\varphi_n(\rho)\approx N_nJ_0(q_n\rho)/\pi^{1/2}l_\perp$, normalized on a disc
of radius $l_\perp$, $N_n=1/|J_1(r_{n+1})|$, $r_{n+1}$ being the $(n+1)$-th root of
$J_0$. Separating from $G_c$ the terms $n\leq n_E$, the series for $n>n_E$ can be 
approximated by an integral over $q$ emerging from the continuum limit $q_n\rightarrow q$
and valid when $r',r\ll l_\perp$. Note that $q$ starts at $q_{1+n_E}>k$. One then compares
real and imaginary parts of $G_c$ in Eq(\ref{gc}) with a suitable expansion of $G_{1,2}$
in {\em cylindrical} coordinates~\cite[Prob.7.9]{morse1953} in Eq(\ref{green-3D}). This 
comparison leads to 
\begin{subequations} 
\label{parameters}
\begin{eqnarray}
\label{gamma}
& & \hspace{-1em}
\gamma_{\pm} = 1/2 \pm\gamma/2\, ,
	\hspace{2em}\gamma\equiv \sum_{n=0}^{n_E}2N_n^2/kk_nl_\perp^2\,,\\ 
\label{deltac}
& & \hspace{-2.4em} \Delta_c(\bm{r},\bm{r}^\prime)\equiv 
	- \frac{1}{4\pi}\int_0^{p_c}dp\, e^{-p|z-z'|} J_0(q\rho)J_0(q\rho')\, ,
\end{eqnarray}
\end{subequations}
with $q=\sqrt{k^2+p^2}$ and $q_{1+n_E}\equiv\sqrt{k^2+p_c^2}\,$. The homogeneous
(Helmholtz) term $\Delta_c$ corrects the Green's function $\gamma_+G_1+\gamma_-G_2$, with   
$\gamma_++\gamma_-=1$, in order to account for the discreteness due to the
confinement. Within the flatness condition, the above approach is valid for arbitrary
$U(\rho)$. It yields an intrinsic connection between the confined and the free space
scattering approaches (see \cite{olshanii1998a} for parabolic confinement). 

In order to obtain the scattering phases $\delta_l$ that are associated with the spherical
symmetry, we expand the incident state in spherical coordinates employing 
$e^{ik_nz}\varphi_n(\rho) = \sum_l i^l(2l+1)\alpha_{nl}j_l(kr)P_l(\cos{\theta})$, with 
$\alpha_{nl}=N_nP_l(k_n/k)/\pi^{1/2}l_\perp$~\cite{morse1953}. Analogously in $\Delta_c$,
the equivalent expansion is given by $e^{-pz}J_0(q\rho)=\sum_l
i^l(2l+1)P_l(ip/k)j_l(kr)P_l(\cos{\theta})$ stemming from an analytic continuation into
the complex $\theta$-plane ($\theta\rightarrow\pi/2-i\theta$). Inserting these
expressions and Eq.(\ref{green}) into Eq.(\ref{int}) and using Eq.(\ref{gamma}) yields,
for $R_V\ll r\ll l_\perp$, 
\begin{eqnarray}
\label{spherical}
\Psi(\bm{r}) &\approx& \sum_l
  i^l(2l+1) \left[\, \alpha_l + \gamma_l(z) - i\gamma kT_l \,\right]j_l(kr)P_l(\cos{\theta}) 
								\nonumber\\
	& &	\hspace{1em} + \sum_l i^l(2l+1) 
			\left[\, kT_l \,\right] n_l(kr)P_l(\cos{\theta})\, ,
\end{eqnarray}
with $\alpha_l=\sum_{n=0}^{n_E}b_n\alpha_{nl}$. Here $4\pi T_l\equiv i^{-l}\int d^3\bm{r}'
[j_l(kr')P_l(\cos{\theta'})] v(r')\Psi(\bm{r}')$ and $4\pi\gamma_l(z)\equiv\int_0^{p_c}dp
\int_{(z)} d^3\bm{r}'P_l(\pm ip/k)e^{\pm pz'}J_0(q\rho')v(r')\Psi(\bm{r}')$. The 
integration over $\bm{r}'$ for $\gamma_l(z)$ is performed in a finite volume $\Omega$
covering the range of $v(r')$. If $z$ is outside $\Omega$, the positive sign refers to a
positive $z$ and vice-versa. Inside $\Omega$, both signs are needed according to whether 
$z\gtrless z'$. Except for this $z$-dependence of $\gamma_l(z)$ in Eq.(\ref{spherical}),
we have now succeeded in representing the total scattering wave function in spherical
coordinates. 

Noteworthy at this point is the fact that $\gamma_l(z)$ accounts for {\em couplings}
between different angular momenta. Indeed, by using $e^{\pm pz'}J_0(q\rho') =\sum_{l'}
i^{l'}(2l'+1)P_{l'}(\mp ip/k)j_{l'}(kr')P_{l'}(\cos{\theta}')$ and the property
$P_{l'}(\mp u)=(-)^{l'}P_{l'}(\pm u)$ in the definition of $\gamma_l(z)$, one gets a
constant $\gamma_l(z)$ if, for each $l$, only $l'$-waves are kept such that 
$l+l'=\mathrm{even}$. The latter condition is also necessary to obtain non-zero matrix
elements $\langle l\left|U(\rho)\right|l'\rangle$ due to the parity symmetry 
$\bm{r}\rightarrow-\bm{r}$. Therefore, a constant $\gamma_l(z)\approx\gamma_l$ arises 
\begin{equation}
\label{gammal}
\gamma_l = \sum_{l'[l]} (2l'+1)P_{ll'}T_{l'}\, , \hspace{1.5em} l=0,1,2,\dots\, , 
\end{equation}
where $P_{ll'} \equiv k\int_0^{p_c/k}du\, P_l(iu)P_{l'}(iu)$ and $l'[l]$ denotes the sum
over even (odd) $l'$ for even (odd) $l$. Eq.(\ref{gammal}) is equivalent to the condition 
that the ``perturbation'' $U(\rho)$ to the free space scattering does not couple even and
odd angular momenta. 

It is now possible to introduce the phase-shifts $\delta_l$. The solution
Eq.(\ref{spherical}) can be written as ($R_V\ll r\ll l_\perp$)  
\begin{subequations}
\label{sphe-boundary}
\begin{eqnarray}
\label{spherical-delta}
& & \hspace{-2.5em} \Psi(\bm{r}) \approx \sum_l
c_l'\left[\cos{\delta_l}\,j_l(kr) - \sin{\delta_l}\,n_l(kr)\right]P_l(\cos{\theta}), \\
\label{constants}
& & \hspace{-1.5em}
c_l' \equiv \frac{(2l+1)(\alpha_l+\gamma_l)\,i^l}{\cos{\delta_l} - i\gamma \sin{\delta_l}}, 
									\hspace{1em}
T_l \equiv \frac{\alpha_l+\gamma_l}{i\gamma k - k\cot{\delta_l}}\, ,
\end{eqnarray}
\end{subequations} 
where the last two relations {\em define} formally $c_l'$ and $\delta_l$. That this
$\delta_l$ is the actual phase-shift can be seen as follows. On one hand,
Eq.(\ref{spherical-delta}) is the (intermediate) asymptotics $R_V\ll r\ll l_\perp$ of the
solution $\Psi(\bm{r})=\sum_lc_l'R_l(r)P_l(\cos{\theta})$ in the region of $V(r)$. On the
other hand, the free-space scattering solution in this region, i.e., not taking into
account the boundary, is just a {\em different superposition}
$\Psi_{3D}(\bm{r})=\sum_lc_lR_lP_l$ with the {\em same} radial part $R_l$. In other words,
the effect of the confinement $U(\rho)$ is to change the superposition coefficients from
$c_l$ to $c_l'$ while keeping the scattering phases of the free-scattering problem. Then
the second relation in Eq.(\ref{constants}) together with Eq.(\ref{gammal}) gives a {\em
matrix equation for $T_l$} in terms of $\delta_l$, i.e., for $l=0,1,2,\dots$ 
\begin{subequations}
\label{main}
\begin{equation}
\label{tmatrix}
\left(i\gamma k - k\cot{\delta_l}\right)T_l = \alpha_l 
	+ \sum_{l'[l]} (2l'+1)P_{ll'}T_{l'}\, .
\end{equation}
Finally, the effective amplitude $f_n^\pm$ is given by expanding $e^{\pm
ik_nz'}\varphi_n(\rho')$ in the integrand of Eq.(\ref{f1d}), thus 
\begin{equation}
\label{f1d-spherical}
f_n^\pm = f_{ng} \pm f_{nu} \equiv
\left( \sum_{l\,\mathrm{even}} \pm \sum_{l\,\mathrm{odd}} \right)
	\frac{(2l+1)4\pi \alpha_{nl}}{2ik_n}T_l\, .
\end{equation}
The relationship between the amplitudes in Eq.(\ref{f1d-spherical}) and the matrix
elements $T_l$ of Eq.(\ref{tmatrix}) constitutes the main result of our formalism. 

{\em Current Conservation}. Inserting Eqs.(\ref{tmatrix},\ref{f1d-spherical}) into
Eq.(\ref{c-boundary}), the probability conservation should follow. From the total current
along the $z$-axis, the conservation condition is  
\begin{equation}
\label{conservation}
\hspace{-0.15em}
0=\sum_{n=0}^{n_E}(|f_{ng}|^2 + \mathrm{Re}\{b_n^\ast f_{ng}\}
		+ |f_{nu}|^2 + \mathrm{Re}\{b_n^\ast f_{nu}\})k_n.
\end{equation}
\end{subequations}
In the remainder of this paper, we analyse the scattering process given by the leading
terms of Eqs.(\ref{main}). We consider first the case of the single mode regime in more
detail, followed by the case of transverse excitations and angular momenta couplings. 

{\em Single Mode Resonances}. When only the ground state ($n_E=0$, $b_n=\delta_{0n}$,
$k^2=q_0^2+k_0^2$) represents an open channel, the symmetric and antisymmetric sectors of 
Eq.(\ref{asymptotic}), $\Psi(\bm{r})=[\psi_g(z)+\psi_u(z)]\,\varphi_0(\rho)$, are given
respectively by (for $z\gtrless 0$) 
\begin{subequations}
\label{sectors}
\begin{eqnarray}
\label{sectors-g}
\hspace{-2em}
\psi_g(z) &=& (1+f_{0g})\cos{(k_0z)}+if_{0g}\sin{(k_0|z|)},\\
\hspace{-2em}
\psi_u(z) &=& i(1+f_{0u})\sin{(k_0z)}\pm f_{0u}\cos{(k_0z)}.
\end{eqnarray}
\end{subequations}
In the context of collisions between identical particles, it is clearly seen that, at
resonance $f_{0g}=-1$, the bosonic sector $\psi_g$ is mapped into a non-interacting
$f_{0u}=0$ pair of (spin-polarized) fermions, the well known fermionization of
impenetrable bosons. Now, the inverse is also seen to occur for $\psi_u$  at the fermionic
resonance, $f_{0u}=-1$, first obtained in~\cite{granger2004a}. A further insight is gained
by setting  
\begin{equation}
\label{sectors-normalized}
f_{0g,u} = - \left[ 1 + i\cot{\delta_{g,u}}\right]^{-1}.
\end{equation}
The conservation condition Eq.(\ref{conservation}) is then fulfilled for real 1D
phase-shifts $\delta_{g,u}$ and one can rewrite
$\psi_g=e^{i\delta_g}\cos{(k_0|z|+\delta_g)}$ and 
$\psi_u=ie^{i\delta_u}\sin{(k_0z\pm\delta_u)}$. Thus at resonance $|\delta_{g,u}|=\pi/2$
and the above discussed boson-fermion and fermion-boson mappings exist also under
longitudinal confinement, e.g., by imposing $\psi_{g,u}(z=l_\parallel)=0$, as numerically 
verified in Ref.~\cite{granger2004a}.

{\em CIR and bound-states}. The resonance $f_{0g}=-1$ can be calculated from a general
potential $V(r)$ by solving Eq.(\ref{tmatrix}) for even $l$. Since $kR_V\sim
R_V/l_\perp\ll 1$, the phase-shifts $\tan{\delta_l}=\tan{\delta_l(k)}\sim k^{2l+1}\sim
1/l_\perp^{2l+1}$ are generally small~\cite{mott1965} for large $l_\perp$. From
Eq.(\ref{tmatrix}), it follows that $l=0$ is the leading contribution and $f_{0g}$ has the
form compatible with Eq.(\ref{sectors-normalized}) 
\begin{subequations}
\label{leading}
\begin{equation} 
\label{swave}
f_0^\pm \approx f_{0g} \approx - \frac{1}
	{1+i\left[-\,\frac{d_\perp^2}{2a}\left(1-aP_{00}\right)\right]k_0 }\,,  
\end{equation}
where $d_\perp\equiv l_\perp/N_0\,$,   
%$\approx \sqrt{8/3\pi^2}\,l_\perp = 0.52\,l_\perp$, 
$P_{00}=p_c$, and $a$ is the 3D $s$-wave scattering length, $k\cot{\delta_0}\approx
-1/a$. This corresponds to solving for $z$ under an effective 1D pseudopotential
$V_{1D}(z)=g_{1D}\delta(z)$, with the coupling strength 
\begin{equation}
\label{gswave}
g_{1D} = 
  \frac{\hbar^2}{\mu} \frac{2a}{d_\perp^2}\left(1-\frac{C'a}{d_\perp}\right)^{-1}, 
		\hspace{1em} C'\equiv d_\perp p_c\, .
\end{equation}
\end{subequations}
As in previous works in the single mode regime (for atom-atom collisions in parabolic
confinement)~\cite{olshanii1998a,bergeman2003,granger2004a}, the resonance
$|g_{1D}|\rightarrow\infty$ at $d_\perp\approx C'a$ requires low longitudinal momenta
$k_0\ll k\sim 1/l_\perp\,$, such that 
$p_c\stackrel{k_0\rightarrow 0}{\longrightarrow}\sqrt{q_1^2-q_0^2}$ 
is not negligible, and large positive scattering length $0<a\sim l_\perp\,$ (meaning that
a weak bound-state of $V(r)$ approaches the threshold~\cite{mott1965}). 
%, yielding $C'\approx \sqrt{20/3}=2.58$.
For scattering by a central field, not only $V(r)$ but also $U(\rho)$ can be quite
general. 

Viewing CIR as a low energy resonant scattering, one could say that bound-states close to
threshold are neither probed at ``high'' energies $k_0\sim 1/l_\perp$ 
($k\rightarrow q_1$, $p_c\rightarrow 0$), nor do they exist for small scattering lengths
($a\ll d_\perp$). However, by calculating the bound-state with energy $E_B'$, this
interpretation for the physical mechanism behind CIR is not accurate: $f_{0g}\approx -1$
occurs before $E_B'$ approaches zero (threshold without confinement), whereas
$E_B'\rightarrow \epsilon_0$ (threshold under confinement) occurs only if $l_\perp$ is
decreased much further below its CIR value. This is explicitly verified e.g. when
$U(\rho)$ is a square-well box of radius $l_\perp\,$: using a cosine approximation to
$J_0$ for its roots, $q_0\approx 3\pi/4l_\perp$ and $q_1\approx 7\pi/4l_\perp$, whence
$C'=d_\perp\sqrt{q_1^2-q_0^2}=\sqrt{20/3}=2.58$ (see~\cite{bergeman2003} for parabolic
$U(\rho)$ and zero-range atom-atom interaction). 

%\begin{figure}
%\includegraphics[scale=0.45]{boundstate}
%\includegraphics[scale=0.55]{bstate} 
%\caption{\label{bstate} Bound-state energy $E_B'$ under confinement, from 
%Eq.(\ref{swave}). As $d_\perp$ is decreased, $E_B'$ (thick-curve) increases from the
%unperturbed $l=0$ state $E_B$, for $d_\perp\gg a\gg R_V$. 
%, and crosses the $E=0$ limit as $d_\perp$ approaches $a$. 
%For $a<0$, the virtual state 
%$E_B$ turns into a real bound-state close to $\epsilon_0$ (thin-line). The crossing of
%$E_B'+(\epsilon_1-\epsilon_0)$ (dashed-line) with the curve $E=\epsilon_0$ and
%the CIR condition almost coincide.}  
%\end{figure}

In fact, the outer $l=0$ bound-state of $V(r)$ in the absence of the confinement has the
energy $E_B\equiv -\,\hbar^2\kappa_B^2/2\mu$ that is related to $a$ via $\kappa_B\approx 
1/|a|$, when $a\gg R_V$~\cite{mott1965}. Under lateral confinement, its tail
$e^{-\kappa_Br}$ is changed to be zero at the edge $r=\rho=l_\perp$. By the uncertainty
principle, this slight squeeze lifts $E_B<0$ by an amount $\epsilon_0$, which can be
sufficient for this state to pass the limit $E=0$ as $l_\perp$ decreases further. 
This new confined bound-state $E_B'$ satisfies Eq.(\ref{diff}) with $k^2$ replaced by 
$2\mu E_B'/\hbar^2$, i.e., $k_0\equiv\pm i\sqrt{q_0^2-2\mu E_B'/\hbar^2}$. Since the 
diverging term $e^{ik_0z}$ should be absent from Eq.(\ref{asymptotic}) and $e^{ik_0|z|}$ 
should decay, $1/f_0^\pm$ must vanish at $\mathrm{Im}\{k_0\}>0$. From Eq.(\ref{swave}), 
for $a<0$, the virtual bound-state with energy $E_B$ turns into a real one with energy 
$E_B'$, which starts at zero for $a/d_\perp=0$ and goes to a positive fraction of 
$\epsilon_0$ as $a/d_\perp\rightarrow -\infty$. This bound-state exists only under
confinement and its experimental measurement is reported in~\cite{moritz2005a}. For $a>0$,
one obtains $E_B'\rightarrow E_B$ for $d_\perp\rightarrow\infty$, as expected. For
$a\rightarrow +\infty$ (or $d_\perp\rightarrow 0$), $E_B'$ tends to a positive fraction of 
$\epsilon_0$. It turns out that the CIR condition (at $a/d_\perp=1/C'=\sqrt{3/20}\approx
0.39$) occurs before $E_B'$ reaches zero (at $a/d_\perp\approx 0.82$). On the other hand,
the CIR almost coincides with the condition $E_B'+(\epsilon_1-\epsilon_0)=\epsilon_0$ (at
$a/d_\perp\approx 0.35$). In Ref.~\cite{bergeman2003}, this last coincidence is exact,
since  $E_B'+(\epsilon_1-\epsilon_0)$ can be associated with a bound-state of the excited 
channels $n\geq 1$ due to a special property of the harmonic oscillator. However, despite
this coincidence, a general mechanism behind CIR needs further study, since 
$E_B'+(\epsilon_1-\epsilon_0)$ has no clear meaning yet beyond parabolic guides and
zero-range pseudopotentials.

%{\em Fermions}. For (spin-polarized) fermionic atoms, the leading contribution to $f_{0u}$
%from Eq.(\ref{tmatrix}) is of $p$-wave type 
%\begin{subequations}
%\label{fermionic-resonance}
%\begin{equation}
%\label{f1}
%f_{0u} \approx -\frac{3[P_1(k_0/k)]^2}
%	{1+i\left[\frac{d_\perp^2}{2}\left(k\cot{\delta_1}+3P_{11}\right)\right]k_0 }\,,
%\end{equation}
%with $P_{11} = -p_c^3/3k^2$. In contrast to bosons, no resonance can occur for $k_0\ll k$
%since then $f_{0u}\rightarrow 0$. The other case compatible with the probability
%conservation Eq.(\ref{sectors-normalized}) is $3[P_1(k_0/k)]^2\sim 1$, i.e., $k_0^2\sim 
%q_0^2/2$. Introducing then the $p$-wave scattering volume~\cite{suno2003a},
%$k\cot{\delta_1}\equiv -1/k^2V_p$, Eq.(\ref{f1}) predicts a resonance at {\em 
%negative} scattering volumes $V_p$, 
%\begin{equation}
%\label{v-critical}
%V_p/d_\perp^3=-1/14.39,
%	\hspace{3em}	k_0\sim q_0/\sqrt{2},
%\end{equation}
%\end{subequations}
%having used $p_c^3d_\perp^3=(71/12)^{3/2}\approx 14.39$ for $k_0^2\sim q_0^2/2$. This
%fermionic CIR was first predicted for parabolic confinement in Ref.~\cite{granger2004a}, 
%where $\psi_u(z)/i(1+f_{0u})$ was obtained via the so-called $K$-matrix
%formalism~\cite{aymar1996a} but restrictions on $k_0$ due to probability conservation
%requirements were not analyzed. 

{\em Excited Channels}. At energies $k^2=q_{n_E}^2+k_{n_E}^2> q_0^2$, the case is more
complex. Keeping only the $l=0$ wave as before, the $n$-th scattering amplitude  
$f_n^\pm$ is 
\begin{equation}
\label{high-energy}
f_n^\pm\approx f_{ng} \approx - \frac{\sum_m b_mN_m/N_n}
 {1 + \sigma_n + i\left[-\frac{d_\perp^2}{2}(-k\cot{\delta_0} - P_{00})\right]k_n},
\end{equation}
where $0\leq m,n\leq n_E$, $P_{00}=(q_{1+n_E}^2-k^2)^{1/2}$ and
in $\sigma_n\equiv\sum_{m\neq n}N_m^2k_n/N_n^2k_m$, $m=n$ is excluded. For the {\em
single} incoming excited channel $n_E$, i.e., $b_n=\delta_{n,n_E}$, the amplitude
$f_{n_Eg}$ does have the form Eq.(\ref{sectors-normalized}) at small $k_{n_E}$. Thus, CIR
at {\em threshold energies} $k\rightarrow q_{n_E}$ can occur when
$-\tan{\delta_0}/k=d_\perp/C'$ as first indicated in Ref.~\cite{granger2004a} for
parabolic confinement. In a more realistic situation of finite temperatures $T$, however,
for a given energy each $b_n$ has the same weight (depending on $E/T$ and with random
phases). Since $f_{ng}=-b_n$ cannot be met for all $n$ simultaneously, one expects no
sharp resonance, with the transmission and reflection probabilities being distributed
among all channels according to~Eq.(\ref{conservation}). 

{\em $l$-couplings}. In the single mode regime, Eq.(\ref{tmatrix}) is also an equation for
$t_l\equiv T_l/k_0$ without the singularity $\gamma\sim k_0^{-1}$. If then $\sum_{l'[l]}
(2l'+1)P_{ll'}t_{l'}$ on the r.h.s converges, one can neglect it compared to 
$\alpha_l$ for $k_0\rightarrow 0$, and $t_l\approx \alpha_l/[i\gamma 
k_0k-(2l+1)k_0P_{ll}-k_0k\cot{\delta_l}]$ is well behaved. Thus, angular momentum
{\em couplings} should be negligible for $k_0\rightarrow 0$ and the series
Eq.(\ref{f1d-spherical}) of individual momenta $l$ is dominated by $l=0$ since
$\delta_l\sim k^{2l+1}\sim 1/l_\perp^{2l+1}$ are small, justifying Eq.(\ref{swave}). 
This does not apply straightforwardly to the excited channel case, whose approximation is
based only on the smallness of $\delta_l\,$.

{\em Discussion}. Consider now the case $U(\rho)=\mu\omega_\perp^2\rho^2/2$ of harmonic
confinement, $\mu$ being the reduced mass. In Eq.(\ref{gswave}), the oscillator length 
$a_\perp\equiv(\hbar/\mu\,\omega_\perp)^{1/2}$ should replace $d_\perp\equiv l_\perp/N_0$ 
instead of $l_\perp\,$. This is due to tunneling, since $|\varphi_n(\rho)|^2\sim  
e^{-\rho^2/a_\perp^2}$ is small at $\rho\approx l_\perp$ (as in the square-well case) only 
if $l_\perp>a_\perp$. Then
$\epsilon_1-\epsilon_0\equiv\hbar^2(q_1^2-q_0^2)/2\mu=2\hbar\omega_\perp$ and
$C'=d_\perp\sqrt{q_1^2-q_0^2}=2\,$. The difference to $C=1.4603\dots$ of
Ref.~\cite{olshanii1998a} originates from the continuum limit in
Eq.(\ref{green-3D}) and Eq.(\ref{deltac}). Indeed, from Eq.(9) of
Ref.~\cite{olshanii1998a}, the continuum approximation for $C$ is
$C\equiv\mathrm{lim}_{s\rightarrow\infty}(\int_0^sds'/\surd{s'}-\sum_{s'=1}^s 1/\surd{s'})
\approx\int_0^1 ds'/\surd{s'}=2\,$. In addition, this comparison reveals the nature of the
``irregular'' part $1/z$ of $\Psi(\bm{r})$ for the pseudopotential approximation (see
Eq.(8) of Ref.~\cite{olshanii1998a} or the equivalent $s$-wave expansion in Eq.(9) of
Ref.~\cite{petrov2000b}). This is the singular part of the free-space Green's function
$\gamma_+G_1+\gamma_-G_2$, with $\gamma_++\gamma_-=1$, and originates from the sum of the
excited transverse levels. As a result, one expects certain details of the guide to be 
unimportant, except for the low lying levels which account for the terms $\gamma$ and
$\Delta_c$ and the bound-state $E_B'$. 

We have provided a general treatment of quantum scattering in confined geometries. For 
scattering by obstacles inside the guide, the treatment should be applicable to a variety 
of central force fields $V(r)$ and confining potentials $U(\rho)$. For ultracold atomic
collisions, non-parabolic guides can be considered with restrictions due to the center of
mass. The 1D scattering amplitude is given in terms of the free-space phase shifts
$\delta_l$ and their couplings among each other. This covers the case of higher energies
and a transversal multi-channel incident state. In the single mode regime, we have shown
that the CIR is closely related to the behaviour of a confined bound state. 
%We note
%finally that CIR in collisions of (spin-polarized) fermions, for which the odd sector
%$f_{nu}$ is involved, are not considered here. 

%\begin{acknowledgments}
The Brazilian Agency CNPq, the German A. v. Humboldt Foundation and the DFG
Schwerpunktprogramm: ``Wechselwirkung in Ultrakalten Atom- und Molek\"ulgasen'' are 
acknowledged for financial support. 
%\end{acknowledgments}

%\bibliography{paper}% Produces the bibliography via BibTeX.

\begin{thebibliography}{99}
\bibitem{intro} Folman R {\it et al.} 2002 {\it Adv. At. Mol. Opt. Phys.} {\bf 48} 263; 
Reichel J 2002 {\it Appl. Phys.} B {\bf 75} 469; 
Grimm R {\it et al.} 2000 {\it Adv. At. Mol. Opt. Phys.} {\bf 42} 95 
\bibitem{morsch2002a} Morsch O {\it et al.} 2002 {\it Phys. Rev.} A {\bf 66} 021601
\bibitem{cataliotti2003a} Cataliotti F S {\em et al.} 2003 
{\it J. Phys. B: At. Mol. Phys.} {\bf 5} S17 
\bibitem{paredes2004a} Paredes B {\it et al.} 2004 {\it Nature} {\bf 429} 277
\bibitem{dettmer2001a} Dettmer S {\it et al.} 2001 {\it Phys. Rev. Lett.} {\bf 87} 160406
\bibitem{petrov2000b} Petrov D S {\it et al.} 2000 {\it Phys. Rev. Lett.} {\bf 84} 2551
\bibitem{diffg} Busch T {\it et al.} 1998 {\it Found. Phys.} {\bf 28} 549; 
Blume D and Greene C H 2002 {\it Phys. Rev.} A {\bf 65} 043613; 
Bolda E L {\it et al.} 2003 {\it Phys. Rev.} A {\bf 68} 032702 
\bibitem{olsson1981} Bostr\"om A and Olsson P 1981 {\it J. Appl. Phys.} {\bf 52} 1187;
Olsson S 1994 {\it Q. J. Mech. Appl. Math.} {\bf 47} 583
\bibitem{olshanii1998a} Olshanii M 1998 {\it Phys. Rev. Lett.} {\bf 81} 938
\bibitem{bergeman2003} Bergeman T {\it et al.} 2003 {\it Phys. Rev. Lett.} {\bf 91} 163201
\bibitem{granger2004a} Granger B E and Blume D 2004 {\it Phys. Rev. Lett.} {\bf 92} 133202
\bibitem{peano2004a} Peano V {\it et al.} 2004 {\it Preprint} cond-mat/0411517
\bibitem{morse1953} Morse P M and Feshbach H 1953 {\it Methods of Theoretical Physics}
(Boston: McGraw-Hill)
\bibitem{mott1965} Mott N F and Massey H S W {\it The Theory of Atomic Collisions}
(Oxford: Oxford University Press)
\bibitem{moritz2005a} Moritz H {\it et al.} 2005 {\it Preprint} cond-mat/0503202
\end{thebibliography}

\end{document}